\documentclass[twocolumn,showkeys,showpacs,preprintnumbers,prd,superscriptaddress,nofootinbib]{revtex4-1}

\usepackage{graphicx}
\usepackage{epsf}
\usepackage{bm}
\usepackage{amsmath}
\usepackage{amsfonts}
\usepackage{amssymb}
\usepackage{epstopdf}
\usepackage{natbib}
\usepackage{hyperref}
\usepackage{color}
\usepackage{verbatim}
\usepackage{multirow}
\usepackage{bm}

\begin{document}

\title{Measurements of $H_0$ in modified gravity theories: The role of lensed quasars in the late-time Universe}

\author{Rocco D'Agostino}
\email{rdagostino@na.infn.it}
\affiliation{Dipartimento di Fisica ``Ettore Pancini'', Universit\`a di Napoli “Federico II”, Via Cinthia, I-80126, Napoli, Italy.}
\affiliation{Istituto Nazionale di Fisica Nucleare (INFN), Sezione di Napoli, Via Cinthia, I-80126, Napoli, Italy.}

\author{Rafael C. Nunes}
\email{rafadcnunes@gmail.com}
\affiliation{Divis\~ao de Astrof\'isica, Instituto Nacional de Pesquisas Espaciais, Avenida dos Astronautas 1758, S\~ao Jos\'e dos Campos, 12227-010, S\~ao Paulo, Brazil}

\begin{abstract}

In this work, we obtain measurements of the Hubble constant in the context of modified gravity theories. We set up our theoretical framework by considering viable cosmological $f(R)$ and $f(T)$ models, and we analyzed them through the use of geometrical data sets obtained in a model-independent way, namely, gravitationally lensed quasars with measured time delays, standard clocks from cosmic chronometers, and standard candles from the Pantheon Supernovae Ia sample. We find $H_0=(72.4\pm 1.4)$ km s$^{-1}$ Mpc$^{-1}$ and $H_0=(71.5\pm 1.3)$ km s$^{-1}$ Mpc$^{-1}$ for the $f(R)$ and $f(T)$ models, respectively. Our results represent 1.9\% and 1.8\% measurements of the Hubble constant, which are fully consistent with the local estimate of $H_0$ by the Hubble Space Telescope. We do not find significant departures from general relativity, as our study shows that the characteristic parameters of the extensions of gravity beyond general relativity are compatible with the $\Lambda$CDM cosmology. Moreover, within the standard cosmological framework, our full joint analysis suggests that it is possible to measure the dark energy equation of state parameter at 1.2\% accuracy, although we find no statistical evidence for deviations from the cosmological constant case.

\end{abstract}



\maketitle
\section{Introduction}

Several astronomical observations predict that the Universe is currently in an accelerated expansion phase \cite{Perlmutter99, Planck16,Haridasu17}. The theoretical modelling that explains such evidence is certainly one of the biggest open problems in contemporary physics and astronomy. Over the last two decades, the Lambda-Cold Dark Matter ($\Lambda$CDM) model has been shown to explain with great precision the observations in the most different scales and cosmic distances. Due to this great success, such a scenario is considered the standard cosmological model.

Nowadays, we have increasingly accurate measurements of the cosmological parameters that challenge the consensus on the $\Lambda$CDM model. Certainly, the most significant tension with the standard model prevision is the observed value of the present cosmic expansion rate, quantified by the Hubble constant, $H_0$. Analyses of the cosmic microwave background (CMB) observations by the Planck Collaboration, assuming the $\Lambda$CDM baseline as input scenario, obtained $H_0=(67.4 \pm 0.5)$ km s$^{-1}$Mpc$^{-1}$ \cite{Planck2018}. On the other hand, model-independent local measurements by the Hubble Space Telescope (HST) showed that $H_0= (74.03 \pm 1.42)$ km s$^{-1}$Mpc$^{-1}$ \cite{R19}, which is in $4.4 \sigma$ tension with Planck's estimate. Moreover, the H0LiCOW Collaboration has revealed its measurement of $H_0$ from its blind (i.e. model-independent) analysis of gravitationally lensed quasars with measured time delays, showing $H_0= (73.3^{+1.7}_{-1.8})$ km s$^{-1}$Mpc$^{-1}$ \cite{H0LiCOW}. This value is in 3.1$\sigma$ tension with the Planck CMB data, increasing to 5.3$\sigma$ when combined with the HST result. Obviously, such a large discrepancy in  the $H_0$ measurements has led to examine the model-dependency of the CMB data or possible underestimated systematic effects in the analysis of the $H_0$ parameter. Therefore, it has been widely discussed in the literature whether a new physics beyond the standard cosmological model can solve the $H_0$ tension (see \cite{H01,H02,H03,H04,H05,H06,H07,H08,H09,H10,H11,H12,H13} for a short list).

Extensions of General Relativity (GR) have been proposed (see \cite{MG_review01,MG_review02, MG_review03,MG_review04} for a review) and exhaustively investigated to explain the observational data at both cosmological and astrophysical levels. The additional gravitational degree(s) of freedom from the modified gravity models quantify extensions of the $\Lambda$CDM cosmology and can drive the accelerating expansion of the Universe at late times. Several of these extensions have shown to fit the data well, leading to a possible theoretical degeneracy\footnote{It has been argued and explored that information from gravitational wave (GW) observations, in particular measures on the propagation speed of GWs, can strongly discriminate among possible extensions of GR. See \cite{GW01,GW02,GW03,GW04,GW05} for discussions in this regard.}. Among viable candidates for modified gravity theories, two classes of theories have been well accepted and investigated in literature, namely, the $f(R)$ gravity and $f(T)$ gravity. 
The $f(R)$ scenarios are gravitational modifications that add higher-order corrections to the Einstein-Hilbert action, extending the Ricci scalar $R$ to an arbitrary function $f(R)$.  We refer to \cite{fR_review} for a review on the $f(R)$ gravity. The $f(R)$ gravity has been tested against several different data and some viable $f(R)$ models have been constrained at different cosmological scales \cite{fR01,fR02,fR03,fR04,fR05,fR06,fR07,fR08,fR09,fR10}. However, one can equally construct the gravitational modifications starting from the torsion-based formulation, and specifically from the Teleparallel Equivalent of General Relativity (TEGR) \cite{TEGR}. In this theory, the Lagrangian is the torsion scalar $T$, and its simplest generalization is represented by the $f(T)$ gravity (see \cite{fT_review} for a review). Also, the $f(T)$ theories have been shown to be a strong and viable modified gravity candidate in alternative to GR \cite{fT01,fT02,fT03,fT04,fT05,fT06,fT07,fT08,fT09,fT10,fT11}.

The main aim of this work is to the use the gravitationally lensed quasars with measured time-delays compiled by the H0LiCOW Collaboration to obtain new observational constraints on both $f(R)$ gravity and $f(T)$ viable models. In particular, these frameworks have proven to be important for measuring $H_0$ parameter with excellent accuracy, as we shall discuss in the following. Hence, it is interesting to check whether alternative gravitational models could provide an explanation to the standing $H_0$ tension.
To do that, we will complement the time-delay distance data with other geometrical probes such as standard candles from type Ia Supernovae (SN Ia), and standard clocks from cosmic chronometers, which are obtained without assuming a cosmological model. Employing these data, we will be able to obtain accurate estimates of the free parameters of the theories, especially the $H_0$ parameter, and check the feasibility of the models. For the quantitative discussion, we will also analyze the $\Lambda$CDM and $w$CDM models in light of the these data and, through a statistical Bayesian comparison, we will interpret the evidence for all the models beyond the $\Lambda$CDM scenario under consideration.

The manuscript is organized as follows. In Sec.~\ref{models}, we provide a brief description of the cosmological dynamics of the $f(R)$ and $f(T)$ gravity theories. In Sec.~\ref{data}, we present the data sets and our methodology to analyze them, whereas in Sec.~\ref{results} we present our main results. In Sec.~\ref{Bayesian}, we statistically compare the predictions of the different theoretical scenarios, and finally, in Sec.~\ref{conclusions}, we summarize our conclusions and indicate the perspectives of our work. 

Throughout the text, we use units such that $c=\hbar=1$, and the notation $\kappa\equiv 8\pi G=M_P^{-2}$, where  $M_P$ is the reduced Planck mass, and $G$ is the gravitational constant. As usual, the symbol dot indicates derivative with respect to the cosmic time, and a subscript zero refers to any quantity evaluated at the present time.

\section{Theoretical framework}
\label{models}
In what follows, we describe in a nutshell the theoretical framework of our study.
\vspace{-0.35cm}
\subsection{$f(R)$ gravity}

We start with a brief review of the $f(R)$ cosmology. The $f(R)$ gravitational theories consist in extending the Einstein-Hilbert action in the form
\begin{equation}
S = \int d^4 x \sqrt{-g}\,\, \dfrac{M_P^2}{2}f(R) + S_m\ ,
\label{action0}
\end{equation}
where $g$ is the determinant of the metric tensor, $f(R)$ is a generic function of the Ricci scalar, and $S_m$ is the action of matter fields. For $f(R) = R$, the GR case is recovered.

Let us now consider a spatially flat FLRW Universe dominated by pressureless matter (baryonic plus dark matter) and radiation with energy densities $\rho_m$, $\rho_r$ and pressures $P_m$, $P_r$, respectively. The modified Friedmann equations in the metric formalism are given by \cite{fR_review}
\begin{eqnarray}
3FH^2=8\pi G  \left(\rho_m+\rho_r\right) , \label{FR1a} \\
-2F\dot{H} = 8\pi G  \left( \rho_m  +\rho_r + P_r \right)+\ddot{F}-H\dot{F}\ ,
\label{FR2a}
\end{eqnarray}
where $F\equiv \frac{\partial f}{\partial R}$. Moreover, one obtains the following useful relation:
\begin{equation}
R=6\left(2H^2+\dot{H}\right).
\end{equation}

In order to move on, we need to specify some $f(R)$ function. Adopting the formalism presented in \cite{Basilakos, Rafael}, one can write
\begin{equation}
\label{fR_models}
f(R) = R - 2 \Lambda y(R, b)\ ,
\end{equation}
where the function $y(R, b)$ quantifies the deviation from Einstein's gravity, i.e. the effect of the $f(R)$ modification, through the parameter $b$. 

We thus consider viable models that have up to two parameters, where the $f(R)$ function is given by Eq.~(\ref{fR_models}). This methodology has been used earlier to investigate the observational constraints on $f(R)$ gravity in \cite{Basilakos, Rafael}. In this respect, one of the most well-known scenarios in the modified gravity theory literature is the Hu-Sawicki (HS) model \cite{Hu:2007nk}, which satisfies all the dynamics conditions required for a given $f(R)$ function. The function $y(R, b)$ for the HS model reads
\begin{equation}
\label{y_HS}
y(R, b) = 1- \dfrac{1}{1+ \Bigl(\dfrac{R}{\Lambda b} \Bigr)^n}\ ,
\end{equation}
where $b>0$ and we assume $n = 1$. We refer to \cite{Basilakos, Rafael} for more details.

\subsection{$f(T)$ gravity}
Inspired by the $f(R)$ extensions of GR, we can generalize $T$ to a function $T + f(T)$, constructing the action of $f(T)$ gravity as \cite{Linder}
\begin{eqnarray}
\label{actionbasic}
S = \frac{1}{16\pi G}\int d^4x\ e \left[T+f(T)\right] + S_m\ ,
\end{eqnarray}
with $e = \text{det}(e_{\mu}^A) = \sqrt{-g}$ \footnote{We use the
vierbein fields $e^\mu_A$, which form an orthonormal base on the tangent space
at each manifold point $x^{\mu}$. The metric then reads
$g_{\mu\nu}=\eta_{A B} e^A_\mu e^B_\nu$.} and where $S_m$ is the action for matter fields. We note that the TEGR is restored when $f(T)=0$, whereas, for $f(T)=const$, we recover GR with a cosmological constant, i.e. the $\Lambda$CDM model.
In the action above, the torsion scalar $T$ is constructed by contractions of the torsion tensor $T^{\rho \mu \nu}$ as \cite{JGPereira}
\begin{equation}
\label{Tscalar}
T\equiv\frac{1}{4}
T^{\rho \mu \nu}
T_{\rho \mu \nu}
+\frac{1}{2}T^{\rho \mu \nu }T_{\nu \mu\rho}
-T_{\rho \mu}{}^{\rho }T^{\nu\mu}{}_{\nu}\ .
\end{equation}
Variation of the action (\ref{actionbasic}) with
respect to the vierbeins provides the field equations:
\begin{eqnarray}
\label{eom}
&&\!\!\!\!\!\!\!\!\!\!\!\!\!\!\!
e^{-1}\partial_{\mu}(ee_A^{\rho}S_{\rho}{}^{\mu\nu})[1+f_{T}]
 +
e_A^{\rho}S_{\rho}{}^{\mu\nu}\partial_{\mu}({T})f_{TT}
\nonumber \\ 
&&-[1+f_{T}]e_{A}^{\lambda}T^{\rho}{}_{\mu\lambda}S_{\rho}{}^{\nu\mu}+\frac{1}{4} e_ { A
} ^ {
\nu
}[T+f({T})] \nonumber \\
&&= 4\pi Ge_{A}^{\rho}
\left[{\mathcal{T}^{(m)}}_{\rho}{}^{\nu}+{\mathcal{T}^{(r)}}_{\rho}{}^{\nu}\right],
\end{eqnarray}
where $f_{T}\equiv \partial f/\partial T$, $f_{TT}\equiv \partial^{2} f/\partial T^{2}$,
while ${\mathcal{T}^{(m)}}_{\rho}{}^{\nu}$ and  ${\mathcal{T}^{(r)}}_{\rho}{}^{\nu}$
are the matter and radiation energy-momentum tensors, respectively.

We then focus on homogeneous and isotropic space-time. Thus, the flat FLRW background metric corresponds to the following
choice for the vierbiens:
\begin{equation}
\label{weproudlyuse}
e_{\mu}^A={\rm
diag}(1,a,a,a)\ ,
\end{equation}
where $a$ is the cosmic scale factor. Inserting the vierbein (\ref{weproudlyuse}) into the field equations
(\ref{eom}), we obtain the Friedmann equations:
\begin{eqnarray}\label{background1}
&&H^2= \frac{8\pi G}{3}(\rho_m+\rho_r)
-\frac{f}{6}+\frac{Tf_T}{3}\ , \\ \label{background2}
&&\dot{H}=-\frac{4\pi G(\rho_m+P_m+\rho_r+P_r)}{1+f_{T}+2Tf_{TT}}\ ,
\end{eqnarray}
where $H\equiv\dot{a}/a$ is the Hubble parameter. In the above relations, we have used the relation
\begin{eqnarray}
\label{TH2}
T=-6H^2,
\end{eqnarray}
which arises straightforwardly from the FLRW metric through Eq.~(\ref{Tscalar}).
Defining the quantity $E\equiv H/H_0$, one can thus rewrite Eq.~(\ref{background1}) as 
\begin{eqnarray}
\label{Mod1Ez}
E^2(z,{\bf r})=\Omega_{m0}(1+z)^3+\Omega_{r0}(1+z)^4+\Omega_{F0} y(z,{\bf r})
\end{eqnarray}
where we have introduced the redshift $z\equiv a^{-1}-1$ and
\begin{equation}
\label{LL}
\Omega_{F0}=1-\Omega_{m0}-\Omega_{r0} \;,
\end{equation}
with $\Omega_{i0}=\frac{8\pi G \rho_{i0}}{3H_0^2}$ being the corresponding
density parameters at present. In this case, the effect of the $f(T)$ modification  is
encoded in the function  $y(z,{\bf r})$ (normalized to
unity at   present time), which depends on $\Omega_{m0},\Omega_{r0}$, and the
$f(T)$-form parameters $r_1,r_2,...$, namely \cite{Nesseris_fT, Rafael02},
\begin{equation}
\label{distortparam}
 y(z,{\bf r})=\frac{1}{T_0\Omega_{F0}}\left[f-2Tf_T\right].
\end{equation}
We note that, due to  (\ref{TH2}), the additional term (\ref{distortparam}) is a function of the Hubble parameter only.

In this work, we consider the parametric form given by the power-law model \cite{Bengochea}
\begin{equation}
\label{modf1}
f(T)=\alpha (-T)^{b}\ ,
\end{equation}
where $\alpha$ and $b$ are the free parameters of the model.
Inserting this $f(T)$ form
into the Friedmann equation (\ref{background1}) evaluated at present, we
find
\begin{eqnarray}
\alpha=(6H_0^2)^{1-b}\frac{\Omega_{F0}}{2b-1}\ ,
\end{eqnarray}
while (\ref{distortparam}) yields
\begin{equation}
\label{yLL}
y(z,b)=E^{2b}(z,b) \;.
\end{equation}
Clearly, for $b=0$  the present scenario reduces to the $\Lambda$CDM cosmology. Finally, we mention that one needs  $b<1$ in order to obtain an accelerating expansion. We refer to \cite{Nesseris_fT, Rafael02} for more details.

\section{Data sets and methodology}
\label{data}

Here, we briefly describe the observational data sets and the statistical methods that we use to explore the parameter space of the modified background dynamics presented above. 

\subsection{H0LiCOW}

A powerful geometric method to measure $H_0$ is offered by the gravitational lensing. The time delay between multiple images, produced by a massive object (lens) and the gravitational potential between a light-emitting source and an observer, can be measured by looking for flux variations that correspond to the same source event. This time delay depends on the mass distribution along the line of sight and in the lensing object, and it represents a complementary and independent approach with respect to the CMB and the distance ladder. Due to their variability and brightness, lensed quasars have been widely used to determine $H_0$ through this method (see   \cite{Sereno14,Kumar15,Bonvin17} and references therein). 
One can calculate the time delay between two images $i$ and $j$ as
\begin{equation}
\Delta t_{ij}=D_{\Delta t}\left[\dfrac{(\bm{\theta}_i-\bm{\beta})^2}{2}-\psi(\bm{\theta}_i)-\dfrac{(\bm{\theta}_j-\bm{\beta})^2}{2}+\psi(\bm{\theta}_j)\right] ,
\end{equation}
where $\bm{\theta}_{i,j}$ are the angular positions of the images, $\bm{\beta}$ is the angular position of the source, and $\psi(\bm{\theta}_{i,j})$ is the lens potentials at the image positions. Here, $D_{\Delta t}$ is the ``time-delay distance'',  which is given by \cite{Treu16}
\begin{equation}
D_{\Delta t}=(1+z_l)\dfrac{D_l D_s}{D_{ls}}\	 , 
\end{equation}
where $z_l$ is the redshift of the lens, while $D_l$, $D_s$ and $D_{ls}$ are the angular diameter distances to the lens, to the source, and between the lens and the source, respectively.
The quantity $D_{\Delta t}$ is highly sensitive to $H_0$, with a weak dependence on other cosmological parameters. 

In the present work, we use the six systems of strongly lensed quasars analyzed by the H0LiCOW Collaboration (we refer to \cite{H0LiCOW} for the details).  The likelihood probability function for the $D_{\Delta t}$ data points reads
\begin{equation}
\mathcal{L}_\text{H0LiCOW}\propto \exp\left\{-\dfrac{1}{2}\sum_{i=1}^{6}\left[\dfrac{D_{\Delta t, i}^{obs}-D_{\Delta t,i}^{th}}{\sigma_{D_{\Delta t},i}}\right]^2\right\} .
\end{equation}

\subsection{Pantheon}

We also take into account the Pantheon sample \cite{Scolnic:2017caz} of 1048 SN Ia in the redshift region $z \in [0.01, 2.3]$, whose distance moduli are standardized through the SALT-2 light-curve fitter (see \cite{Betoule14,rocco_holo} for details). 
As shown in \cite{Riess18}, under the only assumption of a flat Universe, the full Pantheon catalogue can be compressed into six model-independent $E^{-1}(z)$ measurements. 
Therefore, consistently with our theoretical framework assumptions, we use in our analysis these measurements correlated among them according to the covariance matrix $C_{ij}$ given in \cite{Riess18}.
If, from the one hand, SN data by themselves are not able to constrain the value of $H_0$, as this results in being degenerate with the SN absolute magnitude, on the other hand, such a degeneration can be overcome in combination with other cosmological probes, allowing us to obtain tight constraints on the whole set of cosmological parameters underlying a given theoretical scenario.

In the case of the $E(z)^{-1}$ measurements, the likelihood probability function can be written as
\begin{equation}
\mathcal{L}_\text{Pantheon}\propto \exp \left\{-\dfrac{1}{2}\bm{V}^\text{T}C_{ij}^{-1}\bm{V}\right\} ,
\end{equation}
where $\bm{V}=E_{obs}^{-1}-E_{th}^{-1}$ measures the differences between the observed values and the theoretical expectations.
 
 \subsection{Cosmic chronometers}
 
The late expansion history of the Universe can be studied in a model-independent fashion by measuring the age difference of cosmic chronometers (CC), such as old and passively evolving galaxies that act as standard clocks \cite{Jimenez02, Moresco:2016mzx}. From the spectroscopic measurements of the redshifts between pairs of these galaxies and their differential age, one can obtain an estimate of the Hubble parameter through the relation
\begin{equation}
H(z)=-\dfrac{1}{1+z}\dfrac{dz}{dt}\ .
\end{equation}

In our analysis we consider the 31 uncorrelated measurements of $H(z)$ in the redshift range $0< z < 2$ tabulated in \cite{rocco_cheb}. Confronting these values with the corresponding Hubble expansion rates predicted by the theoretical scenarios, one can construct the likelihood function as
\begin{equation}
\mathcal{L}_\text{CC}\propto \exp\left\{-\dfrac{1}{2}\sum_{i=1}^{31}\left[\dfrac{H^{obs}_i-H^{th}_i}{\sigma_{H,i}}\right]^2\right\} .
\end{equation}

\begin{figure*}
\begin{center}
\includegraphics[width=3.3in]{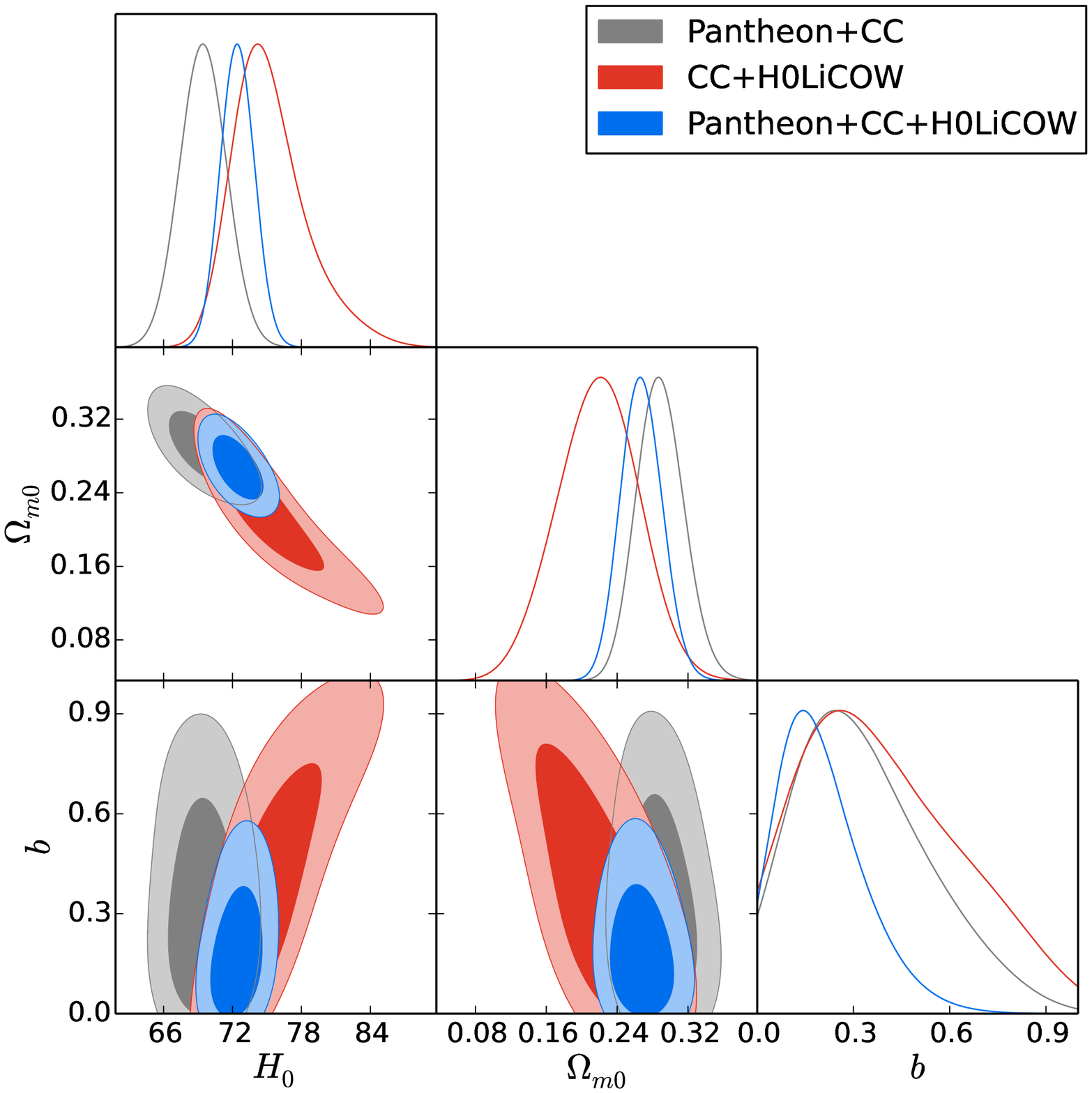} \,\,\,\,
\includegraphics[width=3.3in]{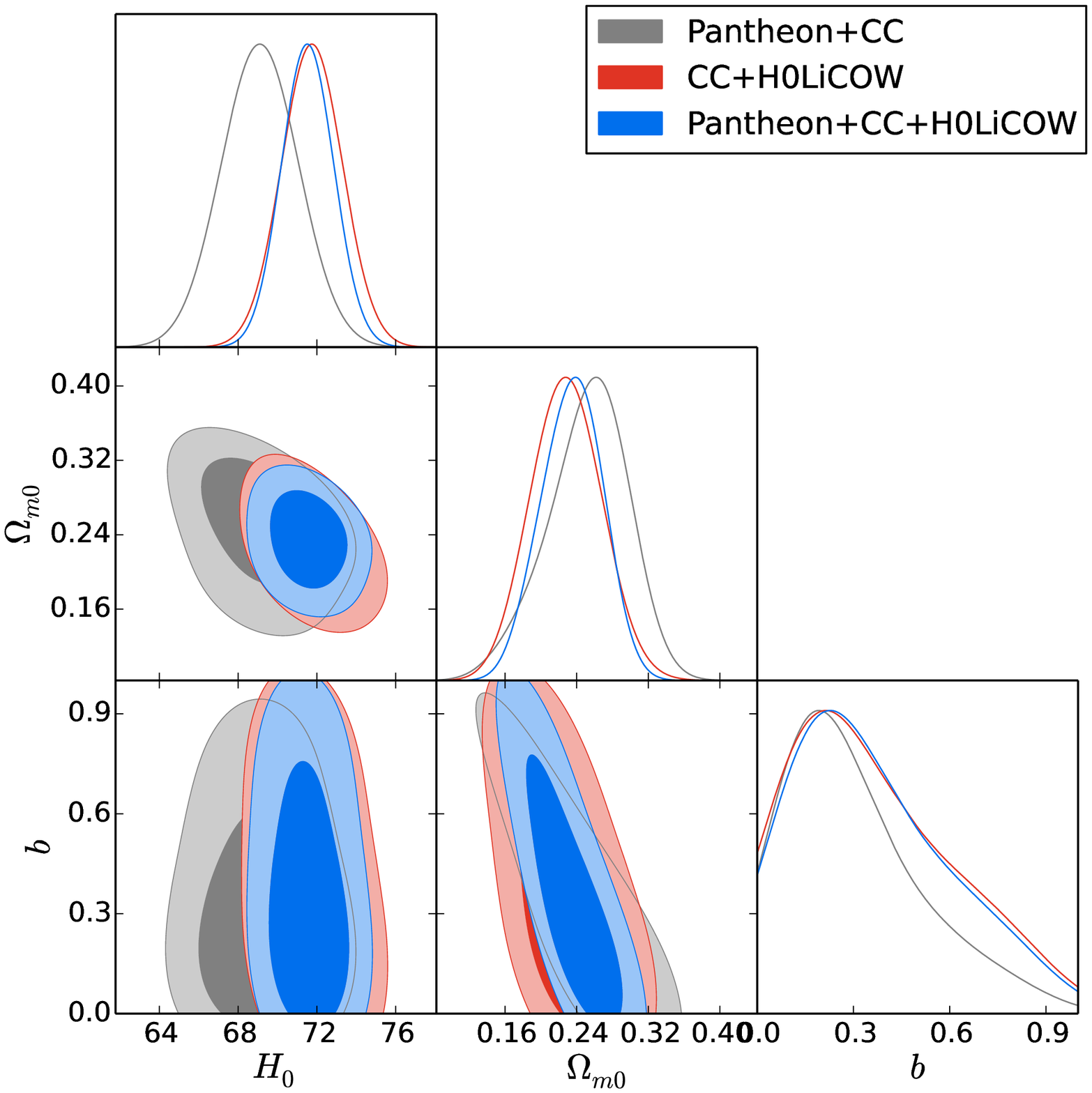}
\caption{Two-dimensional parameter regions and one-dimensional posterior distributions for the Hu-Sawicki $f(R)$ model (left panel) and $f(T)$ power-law model (right panel) as results of the MCMC analysis of different combinations of data.}
\label{fig:fR_fT}
\end{center}
\end{figure*}

\begin{table*}
\begin{center}
\setlength{\tabcolsep}{1em}
\renewcommand{\arraystretch}{2}
\begin{tabular}{c c c c c}
\hline
\hline
Model & Data & $H_0$ & $\Omega_{m0}$ & $b$ \\
\hline
\multirow{3}{*}{$f(R)$}  & CC+Pantheon & $69.5\pm 2.0(3.9)$ & $0.289^{+0.025(0.053)}_{-0.028(0.048)}   $ & $0.32^{+0.17(0.45)}_{-0.25(0.32)}      $\\
&CC+H0LiCOW & $75.2^{+2.4(7.2)}_{-3.7(5.8)}  $ & $0.218^{+0.045(0.086)}_{-0.046(0.090)} $ & $0.37^{+0.22(0.54)}_{-0.32(0.37)}           $\\
&  CC+Pantheon+H0LiCOW & $72.4^{+1.4(2.8)}_{-1.4(2.7)}  $  & $0.267^{+0.023(0.045)}_{-0.023(0.042)}   $  & $0.19^{+0.10(0.29)}_{-0.16(0.19)}      $ \\ 
\hline
\multirow{3}{*}{$f(T)$} & CC+Pantheon & $69.1^{+1.9(3.8)}_{-1.9(3.7)} $  & $0.251^{+0.050(0.084)}_{-0.040(0.094)}  $    & $0.30^{+0.16(0.49)}_{-0.27(0.30)}      $  \\
& CC+H0LiCOW & $71.8\pm 1.5(3.0) $ & $0.228^{+0.039(0.077)}_{-0.039(0.074)}   $  & $0.34^{+0.21(0.56)}_{-0.34(0.34)}      $\\
&  CC+Pantheon+H0LiCOW & $71.5^{+1.3(2.6)}_{-1.3(2.5)}  $  & $0.233^{+0.044(0.072)}_{-0.033(0.083)}   $ &   $0.27^{+0.16(0.49)}_{-0.27(0.27)}    $  \\ 
\hline
\hline
\end{tabular}
\caption{68\% (95\%) C.L. constraints on the Hu-Sawicki $f(R)$ model  and the $f(T)$ power-law model from different combinations of data. $H_0$ is measured in units of km s${}^{-1}$ Mpc${}^{-1}$.}
\label{tab:results modified gravity}
\end{center}
\end{table*}

\subsection{Monte Carlo method}

We perform a statistical analysis of the data sets presented above through a Markov chain Monte Carlo (MCMC) method, based on the Metropolis-Hastings algorithm \cite{Metropolis-Hastings}. Specifically, we analyze the HS $f(R)$ model and the $f(T)$ power-law  model by assuming the following flat priors on the cosmological parameters: $H_0\in[50,90]$ km s${}^{-1}$ Mpc${}^{-1}$, $\Omega_{m0}\in[0,1]$ and $b\in[0,1]$. In our study, we neglect the late-time contribution of radiation $(\Omega_{r0}\approx 0)$.
Moreover, for comparison, we also consider the  standard $\Lambda$CDM model and its one-parameter extension, namely the $w$CDM model, characterized by a constant equation of state parameter for the dark energy fluid ($w$). In this case, we assume the flat prior $w\in [-2.0,-0.3]$.

Our analysis consists in two steps. We first combine the Pantheon + CC data to constrain the cosmological parameters of the theoretical scenarios under consideration, and we then compare these results with the outcomes of the full joint likelihood analysis\footnote{The full joint likelihood is obtained as the product of the individual likelihoods: $\mathcal{L}_\text{joint}=\mathcal{L}_\text{Pantheon}\times \mathcal{L}_\text{CC}\times \mathcal{L}_\text{H0LiCOW}$.} (Pantheon + CC + H0LiCOW), to check the effects of the time-delay quasars measurements on the $H_0$ value. 
Taken individually, the Pantheon and CC data will weakly constrain the full parametric space of the models, especially $H_0$, making this parameter degenerate. These data are not in tension with each other. Thus, we shall consider CC in specific combinations with Pantheon and H0LiCOW.

In this work, we choose not to use baryon acoustic oscillations (BAO) data and CMB distance priors, as our main focus lies on late-time cosmology, and to avoid any possible physical bias (some input fiducial cosmology) towards the standard model. It is worth reminding the reader that the most common BAO measurements considered in the literature are obtained adopting a fiducial cosmology, usually fixed to GR + $\Lambda$, although efforts have being made to analyze BAO data in some model-independent way \cite{bao1,bao2,bao3}. The BAO data are usually used in joint analyses to break a given degeneracy in the parametric space. We note that the minimal combination CC + Pantheon and CC +H0LiCOW are enough to obtain reasonably accurate constraints on our parameters baseline. 
With regard to distance priors from the CMB compressed likelihood, these data are as well model-dependent, as the CMB shift parameter, around some fiducial model, usually the $\Lambda$CDM cosmology\footnote{See \cite{cmb_prior1} for a recent release from the Planck Collaboration, and also \cite{cmb_prior2} for a discussion on model-dependence in CMB distance priors.}.  On the other hand, as also argued in \cite{planck_DE}, the CMB distance priors data should not be used for models with low sound speed or modifications of gravity.

We present our main results in what follows.

\begin{figure}
\begin{center}
\includegraphics[width=3.3in]{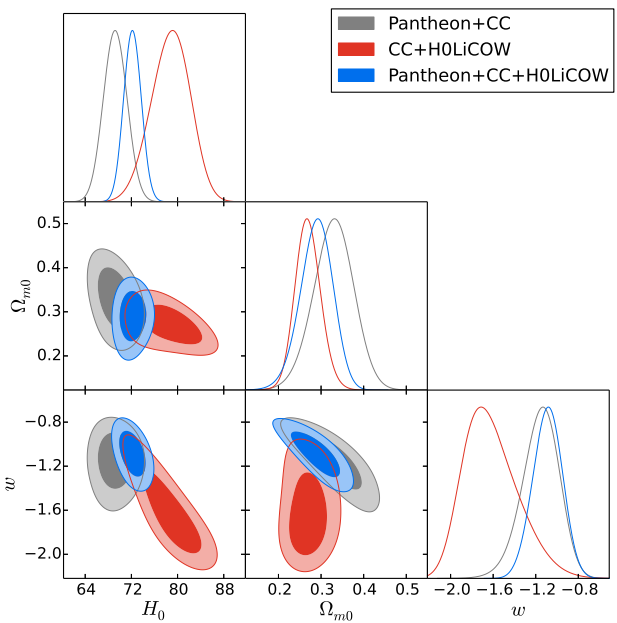}
\caption{Two-dimensional parameter regions and one-dimensional posterior distributions for the $w$CDM  model as results of the MCMC analysis of different combinations of data.}
\label{fig:standard models}
\end{center}
\end{figure}

\section{Results}
\label{results}

\begin{table*}
\begin{center}
\setlength{\tabcolsep}{1em}
\renewcommand{\arraystretch}{2}
\begin{tabular}{c c c c c}
\hline
\hline
Model & Data & $H_0$ & $\Omega_{m0}$ & $w$ \\
\hline
\multirow{3}{*}{$\Lambda$CDM}  & CC + Pantheon & $69.2\pm 1.9(3.7)$ & $0.296^{+0.026(0.056)}_{-0.029(0.051)}   $ &  $-1$\\
& CC+H0LiCOW & $72.3^{+1.5(2.9)}_{-1.5(2.8)}$ & $ 0.256^{+0.031(0.067)}_{-0.034(0.061)}   $ & $-1$\\
&  CC+Pantheon+H0LiCOW & $71.8\pm 1.3(2.5)$  &  $0.272^{+0.021(0.046)}_{-0.023(0.043)}$ & $-1$ \\ 
\hline
\multirow{3}{*}{$w$CDM} & CC+Pantheon & $69.2^{+2.0(3.9)}_{-2.0(3.8)}$ & $0.329^{+0.045(0.087)}_{-0.045(0.094)}$& $-1.15^{+0.18(0.33)}_{-0.16(0.35)} $ \\
& CC+H0LiCOW  & $78.8^{+3.5(6.0)}_{-3.2(6.7)}$ & $0.269^{+0.028(0.060)}_{-0.031(0.054)}   $  & $-1.63^{+0.20(0.53)}_{-0.29(0.45)}     $	\\
&  CC+Pantheon+H0LiCOW  &  $72.2^{+1.5(2.9)}_{-1.5(2.8)}$ & $0.289^{+0.040(0.073)}_{-0.035(0.077)}   $ & $-1.09^{+0.13(0.26)}_{-0.13(0.27)}$   \\ 
\hline
\hline
\end{tabular}
\caption{68\% (95\%) C.L. constraints on the $\Lambda$CDM  and $w$CDM models from different combinations of data. $H_0$ is measured in units of km s${}^{-1}$ Mpc${}^{-1}$.}
\label{tab:results standard models}
\end{center}
\end{table*}
In this section, we present our main results on the cosmological scenarios previously introduced, using different data combinations.  
We note that, in principle, one could choose other parametric $f(R)$ and $f(T)$ functions, but significant differences among parametric models should only have impact when analyzed at the perturbation level.
Since the data analyzed here are all from geometrical origin, different functions should in fact not change the main results on the modified gravity scenarios. Therefore, without loss of generality, we focus on the most viable and studied models in the literature, which have been described in the previous sections, taking into account only geometrical data sets obtained in a model-independent way at low $z$. As we shall see, the new constraints obtained here are competitive in precision on the full parametric baseline of the models under consideration. Also, to our knowledge, this is the first study in which the H0LiCOW data compilation is used to analyze the modified gravity scenarios described above.

In Table \ref{tab:results modified gravity}, we summarize the main results from the statistical analyses of the $f(R)$ gravity and $f(T)$ gravity models. For comparison, in Table \ref{tab:results standard models}, we also show the results concerning  the $\Lambda$CDM and $w$CDM models.
For the $f(R)$ gravity, we find $H_0 = (69.5 \pm 2.0)$ km s${}^{-1}$ Mpc${}^{-1}$, $H_0 = (75.2^{+2.4}_{-3.7})$ km s${}^{-1}$ Mpc${}^{-1}$ and $H_0 = (72.4 \pm 1.4)$ km s${}^{-1}$ Mpc${}^{-1}$ at the 68\% confidence level (C.L.) from CC + Pantheon, CC + H0LiCOW  and Pantheon + CC + H0LiCOW data, respectively. These estimates represent $\sim$2.8\% (CC + Pantheon), $\sim$4\% (CC + H0LiCOW) and $\sim$1.9\% (CC + Pantheon + H0LiCOW) precision measurements. Although all the $H_0$ measurements are compatible with each other, we can see how the Pantheon data influence the results. Combining Pantheon together with CC data tends to generate lower $H_0$ values (compared to CC + H0LiCOW estimates). Since it is expected  that the constraints from H0LiCOW provide high bounds on $H_0$, the inclusion of Pantheon data will reduce the upper boundary on $H_0$, as well as provide strong restrictions on $\Omega_{m0}$\footnote{Constraints from Pantheon data tends to keep the total matter density around $\Omega_{m0} \sim 0.30$ \cite{Riess18}.}. 
As the parametric space $(\Omega_{m0} - H_0)$ is anti-correlated, the joint analysis without Pantheon data, i.e., CC + H0LiCOW, will generate a lower $\Omega_{m}$ value, while $H_0$ tends to have a higher value. Therefore, the influence of the Pantheon data in our joint analysis reflects in bounding $\Omega_{m0}$ towards $\sim 0.30 $ and improving the upper limits in $H_0$ towards lower value with respect to the predictions of H0LiCOW data. The parametric space $(\Omega_{m0} - H_0)$ in the left panel of Fig.~\ref{fig:fR_fT} summarizes this information.

The local measurement obtained by Riess et al. \cite{R19} from observations of long-period Cepheids in the Large Magellanic Cloud (LMC) is $H_0 = (74.03 \pm 1.42$) km s${}^{-1}$ Mpc${}^{-1}$. Thus, the result of our full joint analysis is in full agreement with the local measurement of $H_0$, and in tension at $3.4\sigma$ with the most recent CMB estimate from Planck \cite{Planck2018}.
Regarding to possible deviations from the standard cosmological model, we find the upper limits $b < 0.77 \,, < 0.90 \,$ and $< 0.48$ at the 95\% C.L. from  CC + Pantheon, CC + H0LiCOW and CC + Pantheon + H0LiCOW, respectively. Therefore, our full joint analysis produces a significant improvement in the constraints of the additional parameter of the theory that quantifies deviations from the $\Lambda$CDM cosmology. 
In the left panel of Fig.~\ref{fig:fR_fT}, we show the parameter space of the $f(R)$ model at the 68\% and 95\% C.L. In particular, focussing on the $(b - H_0)$ plane, we can see that these parameters are not strongly correlated. Similar considerations apply also to the $(b - \Omega_{m0})$ plane.

As far as the $f(T)$ gravity is concerned, at the 68\% C.L. we find $H_0 = (69.1 \pm 1.9)$ km s${}^{-1}$ Mpc${}^{-1}$, $H_0 = (71.8 \pm 1.5)$ km s${}^{-1}$ Mpc${}^{-1}$  and $H_0 = (71.5 \pm 1.3)$ km s${}^{-1}$ Mpc${}^{-1}$ from CC+Pantheon, CC+H0LiCOW and CC + Pantheon + H0LiCOW, which represent $\sim$2.7\%, $\sim$2\%  and $\sim$1.8\% precision estimates, respectively. The result of our full joint analysis is almost $3\sigma$ away from the CMB estimate. As also observed in $f(R)$ gravity, within the $f(T)$ gravity framework we note that $b$ is fully compatible with GR at a larger statistical significance. On the other hand, in light of the joint analysis, the constraints on $b$ and $H_0$ are not improved in the same efficient way as in $f(R)$ gravity. The right panel of Fig.~\ref{fig:fR_fT} shows the parameter space of the $f(T)$ model at the 68\% and 95\% C.L. In this case, we can see an anti-correlation in the plane $(b - \Omega_{m0})$ and no correlation in the plane $(H_0 - \Omega_{m0})$. 
We also note less (dark) matter density at late times with respect to the amount predicted by the $\Lambda$CDM cosmology (cf. Table \ref{tab:results standard models}). This is due the fact that the effective dark energy induced via $f(T)$ gravity starts dominating the expansion of the Universe earlier than what predicted by the cosmological constant\footnote{See, for example, Fig.~1 in \cite{Rafael02}.}. More specifically, within the $\Lambda$CDM model, if we assume a spatially flat Universe, i.e., the normalization condition $\Omega_{m} + \Omega_{\Lambda} = 1$ at late times, we can write the transition from deceleration to acceleration phases at a transition redshift $z_t$ as $z_t = [\frac{2(1-\Omega_{m0})}{\Omega_{m0} }]^{1/3} -1$. Assuming the standard value $\Omega_{m0} \approx 0.30$, we have $z_t \approx 0.67$. For reasonable values of the parameter $b$, this transition must happen at $z_t \sim 0.8$ in $f(T)$ gravity. Since, in $f(T)$ gravity, more effective dark energy density compared to $\Lambda$ is predicted, inducing a greater $z_t$ via the relationship $\Omega_{\rm dark \,\, matter} + \Omega_{\rm dark \,\, energy} = 1$, we have less $\Omega_{m}$ at late times. Also, we note that the parametric space $(\Omega_{m0} - H_0)$ is anti-correlated. As the Universe expands faster in $f(T)$ gravity, compared to $\Lambda$CDM, this will generate a greater $H(z=0)$ value in $f(T)$ gravity, and due the anti-correlation with $\Omega_{m0}$, consequently we will have less dark matter density at $z =0$. Contrary to what happens in the case of $f(R)$ gravity, adding  Pantheon to CC + H0LiCOW data does not influence significantly the bounds of the full baseline parameters in $f(T)$ gravity (see right panel of Fig.~\ref{fig:fR_fT}).

We note that, in both modified gravity theories, there is a shift in the value of $H_0$ when lensing is used with respect to the case of only Pantheon + CC. As argued in \cite{H0LiCOW}, the time-delay distance is primarily sensitive to $H_0$, although there is a weak dependence on other parameters, and this cosmographic test can improve the precision of the other probes, demonstrating the strong complementarity. Since analyzing all lenses in a flat $\Lambda$CDM cosmology leads to $H_0$ living in the range $(71.5 - 75.0)$ km s${}^{-1}$ Mpc${}^{-1}$ at the 68\% C.L., it is expected that, when combining H0LiCOW with other cosmological probes not in tension with H0LiCOW data, like Pantheon + CC, the final joint analysis will have natural shift in direction to high $H_0$ values also for models beyond the standard scenario, similarly to what happens within the $\Lambda$CDM cosmology.

The H0LiCOW Collaboration \cite{H0LiCOW} reported  $H_0 = 73.6^{+1.6}_{-1.8}$ km s${}^{-1}$ Mpc${}^{-1}$ and $H_0 = 74.9^{+2.2}_{-2.4}$ km s${}^{-1}$ Mpc${}^{-1}$ using Pantheon + H0LiCOW data for the $\Lambda$CDM and $w$CDM models, respectively. For a direct comparison, we added the CC data in our analysis and we note that these constraints can be improved (see Table \ref{tab:results standard models}). With regard to the dark energy equation of state, we do not find any significant deviations from $w=-1$ from the CC+Pantheon+H0LiCOW analysis, within which $w$ is measured at 1.2\% accuracy and $H_0$ is measured at 2\% accuracy. In the CC+H0LiCOW analysis, we note a predominance for a phantom behavior at the 95\% C.L. (see Fig. 2). This is due to the characteristic of H0LiCOW data to prefer higher $H_0$ values and to the significant anti-correlation of the parametric space $w-H_0$. Including Pantheon data makes $\Omega_{m0}$ constrained around $0.30$ and also improves the bounds on $H_0$. 
We finally note that all the $H_0$ measurements from our full joint analysis, in all scenarios, are fully compatible with each other.

\section{Bayesian evidence}
\label{Bayesian}
\begin{table}
\begin{center}
\setlength{\tabcolsep}{1.2em}
\renewcommand{\arraystretch}{2}
\begin{tabular}{c c c c c}
\hline
\hline
Model &  $\ln(B_{ij})$ & $\Delta$AIC & $\Delta$BIC  \\
\hline
$f(R)$ & 0.693 & 2.00  & 3.76 \\  
$f(T)$ &  0.765 &  2.01 & 3.77  \\  
$w$CDM  & 1.21 & 1.61 & 3.37 \\ 
\hline
\hline
\end{tabular}
\caption{The logarithmic Bayes factor, AIC and BIC  for different cosmological models from the full joint analysis Pantheon+CC+H0LiCOW, calculated with respect to the standard $\Lambda$CDM model.}
\label{tab:statistics}
\end{center}
\end{table}
Here, we perform a statistical comparison of the different cosmological models using the Bayesian evidence (see \cite{Trotta08} for a comprehensive review).
 The posterior probability for a model $M$ described by a set of parameters $\theta$, given the data $D$, is expressed as
\begin{equation}
P(\theta|D,M)=\dfrac{P(D|\theta,M)P(\theta|M)}{P(D|M)} \ ,
\label{eq:Bayes}
\end{equation}
where $P(D|\theta,M)$ corresponds to the likelihood distribution, and $P(\theta|M)$ is the prior probability for $\theta$, assuming the model to hold true. The Bayesian evidence, i.e. the probability of the data given the model, is obtained by integrating Eq.~(\ref{eq:Bayes}) over $\theta$:
\begin{equation}
P(D|M)=\int d\theta\  P(D|\theta,M)P(\theta|M)\ .
\end{equation}
The comparison between two models $M_i$ and $M_j$ is obtained through the ratio of their evidences, $B_{ij}\equiv P(D|M_i)/P(D|M_j)$, known as the \emph{Bayes factor}. This quantity can be thought as the mathematical implementation of Occam's razor, as the average likelihood (predictability) of complex models is lower compared to that of simpler models with a fewer number of parameters.
For our purposes, we can write the Bayes factor as
\begin{equation}
B_{ij}\equiv \dfrac{L(M_i)}{L(M_j)}\ ,
\label{eq:B}
\end{equation}
where $L(M_i)$ is the probability $P(D|M_i)$ to obtain the data $D$ if the model $M_i$ is assumed to be true:
\begin{equation}
\label{L}
L(M_i)=\int d\theta\ \mathcal{L}_i(\theta) p(\theta|M_i) \ ,
\end{equation}
where $\mathcal{L}_i(\theta)=e^{-\chi^2(\theta)/2}$  is the likelihood for the parameter $\theta$, and $p(\theta|M_i)$ is the prior probability  for  $\theta$ within the model $M_i$.

Let us focus on the case of flat priors, as the ones assumed in our MCMC analysis. Specifically, we consider  a cosmological model $M$, described by a set of parameters $\theta\equiv(\theta_1,\dots, \theta_N)$,  each of them assumed to lie in some range $[\theta_n,\theta_n+\Delta\theta_n]_{n=1,\dots,N}$, with no further prior information. Thus, $p(\theta_n|M_i)=1/\Delta\theta_n$, and Eq.~(\ref{L}) can be simply written as\footnote{We refer the reader to \cite{Nesseris12} for further details regarding  Gaussian and flat priors.}
\begin{equation}
L(M)= \left(\prod_{n=1}^N\dfrac{1}{\Delta\theta_n}\right)\int_{\theta_n}^{\theta_n+\Delta\theta_n}\prod_{n=1}^Nd\theta_n\ e^{-\chi^2(\theta)/2}\ .
\label{eq:integral}
\end{equation}
Hence, using the above expression in Eq.~(\ref{eq:B}), one can calculate the Bayesian evidence for the model $M_i$ against the model $M_j$. In general, such a calculation may be computationally demanding and time-consuming in the case of complex likelihoods, and this often requires the use of semi-analytical approximations (see, e.g., \cite{Nesseris12}). In our case, however, the use of only geometrical data and the few number of free parameters of the models simplify this procedure, so that we were able to smoothly compute the integral in Eq.~(\ref{eq:integral}) through the numerical routines in \textit{Mathematica}. 

The interpretation of the Bayes factor is provided by Jeffreys' scale \cite{Jeffreys61}, which can be summarized as follows: if $1<B_{ij}<3$, there is evidence in favour of the model $M_i$ over the model $M_j$, but it is worth only a bare mention; when $3<B_{ij}<20$, the evidence against $M_j$ is definite although not strong; if $20<B_{ij}<150$, this evidence becomes strong, and for $B_{ij}>150$ it is very strong. In our notation $M_j$ refers to $\Lambda$CDM model and $M_i$ refers to the extended scenarios.

However, very often it is useful and convenient to employ alternative methods based on information theory, which represent fair approximations of the Bayesian evidence under specific assumptions aimed at replacing the different prior volumes with a penalty term taking into account the model complexity.

In this respect, the most widely adopted approaches are the Akaike information criterion (AIC) \cite{AIC} and Bayesian information criterion (BIC) \cite{BIC}. 
The AIC is defined through the relation
\begin{equation}
\text{AIC} \equiv -2 \ln  \mathcal{L}_{\rm max} + 2 N  = \chi_{\rm min}^2 + 2N\ ,
 \end{equation}
where $ \mathcal{L}_{\rm max}$ is the maximum likelihood value, and $N$ is the total number of free parameters in the model. 
The AIC is derived from an approximate minimization of the the Kullback-Leibler divergence between the distribution fitted to the data and true model distribution.
For the statistical comparison, the AIC difference between the model under study and the reference model is calculated. This difference in AIC values can be interpreted as the evidence in favour of the model under study over the reference model. It has been argued in \cite{Tan12} that one model can be preferred with respect to another if the AIC difference between the two models is greater than a threshold value $\Delta_{\rm threshold}$. As a rule of thumb, $\Delta_{\rm threshold} = 5$ can be considered the minimum value to assert a strong support in favour of the model with a smaller AIC value, regardless of the properties of the models under comparison \cite{Liddle07}. 
The BIC is defined as
\begin{equation}
\text{BIC} \equiv -2 \ln  \mathcal{L}_{\rm max} + N \ln(k) = \chi_{\rm min}^2 + N \ln(k)\ ,
 \end{equation}
which provides a more severe penalisation against the model with a larger number of free parameters, due to the presence of the logarithm of the total number of data points $(k)$.
The BIC is obtained from a Gaussian approximation of the Bayesian evidence for a large sample size.
The strength of the evidence against the model with higher BIC value can be summarized as follows:  for $0 \leq \Delta\text{BIC} < 2$, there is not enough evidence; for $2\leq \Delta\text{BIC} \leq 6$, there exists a moderate evidence; for $\Delta\text{BIC}> 6$, there is a strong evidence.

Thus, we compare the $f(R)$ model, the $f(T)$ model and the $w$CDM model to the $\Lambda$CDM scenario, which we choose to be the reference model, since it represents statistically the simplest cosmological model with the least number of parameters.
Table \ref{tab:statistics} summarizes our results for the Bayes factor, as well as for the AIC and BIC, for different cosmological models from our full joint analysis. Due to a minimal number of free parameters, the $\Lambda$CDM cosmology is the statistically preferred scenario to best-fit the data, whereas there is weak-to-moderate evidence against the alternative scenarios.
Thus, we found no significant support for deviations from GR.

\section{Conclusions}
\label{conclusions}

Using geometric model-independent low and intermediate redshift data, we obtained measurements of the Hubble constant in the context of modified background dynamics beyond GR. At the same time, we found new constraints on the free parameters of such theories. Particular attention was given to cosmologically viable $f(R)$ and $f(T)$ gravity models, for which we showed that $H_0$ can be measured with an accuracy of 1.9\% and 1.8\%, respectively. Including the time-delays observations from strong gravitationally lensed quasars in our Monte Carlo statistical analysis, our results appear consistent with the local (direct) measurement of $H_0$ from the LMC Cepheid standards, while they are $\gtrsim 3\sigma$ in tension with the CMB estimate based on the $\Lambda$CDM cosmology.

For comparison with previous results found in the literature, when (quasi) model-independent data were used in the context of modified gravity scenarios, $H_0$ was measured at $\sim$5.2\% accuracy in $f(R)$ gravity \cite{Rafael}, and at $\sim$2.4\% and $\sim$2.6\% accuracies in $f(T)$ gravity in \cite{Rafael02} and \cite{fT10}, respectively. Furthermore, analyses containing the full CMB likelihood (temperature, polarization and lensing) in the $f(T)$ gravity context, led to a $\sim$2.8\% accuracy estimate of $H_0$ \cite{H04}. It has also been discussed that, in future analyses with information from gravitational waves, it will be possible to measure $H_0$ at $\sim$1\% accuracy in $f(R)$ gravity \cite{fR08}.

Although the free parameters of the theories analyzed here are constrained in a precise and robust way, we found no significant deviations from GR, and the dynamics of the Universe is compatible with that of the $\Lambda$CDM model at the background level.

Finally, it would be interesting to implement a cosmographic analysis of the time-delay quasars measurements and obtain model-independent constraints on kinematic parameters, using machine learning methods in order to verify their compatibility with the predictions of a given theoretical scenario.

\begin{acknowledgments}
\noindent 
The authors would like to thank the anonymous referee for useful comments and instructive discussions. The authors are also grateful to Alexander Bonilla for useful discussions on Bayesian evidence. R.D. acknowledges the support of INFN (iniziativa specifica QGSKY).
R.C.N. would like to thank the agency FAPESP for financial support under the Project No. 2018/18036-5.  
\end{acknowledgments}


\end{document}